\documentclass[11pt]{article}
\usepackage[left=2.5cm,top=2.5cm,right=2cm,bottom=2.5cm]{geometry}
\usepackage{amsmath, amssymb}

\usepackage{graphicx,subfigure,epsfig,epsf,color}
\usepackage[utf8x]{inputenc}
\usepackage{slashed}
\usepackage{feynmf}
\usepackage{graphicx}
\usepackage{graphics}
\usepackage{epstopdf}
\usepackage{subfigure}
\usepackage{amsfonts}
\usepackage{sectsty}
\usepackage{hyperref}
\usepackage{cite}
\usepackage{pdflscape}
\usepackage{lipsum}
\begin{document}
 \date{}
\title{Emergent Cosmology in Models of Nonlinear Electrodynamics }
\maketitle
 \begin{center}
\author{Payel Sarkar}\footnote{p20170444@goa.bits-pilani.ac.in}~Prasanta Kumar Das\footnote{pdas@goa.bits-pilani.ac.in} \\
 \end{center}
 
 \begin{center}
 Birla Institute of Technology and Science-Pilani, K. K. Birla Goa campus, NH-17B, Zuarinagar, Goa-403726, India
 \end{center}
 \vspace*{0.25in}
 
  \abstract{Nonlinear electrodynamics, which acts as a source of gravity Einstein field equations, leads to emergent cosmology, an alternative solution which can avoid Big Bang singularity. In this paper, we explore the emerging universe in models of non-linear electrodynamics (described by dimensional parameter $\beta$) by using the equation of state parameter $\omega$ and see how the parameter $\beta$ helps the universe to cause a transition from a quasi-static Minkowski phase to the inflationary phase of expansion through the point of emergence and subsequently to the phase of normal thermal expansion. We predict the spectral index parameter $n_s = 0.97467$ (scalar spectral index), $r =0.10133$ (tensor to scalar ratio) and $n_T = -0.01267$ (tensor spectral index) of the inflationary perturbation in emergent cosmology of nonlinear electrodynamics corresponding to $\beta$ = 0.1 and $B_0=10^{-10}$G.}   
  
 \maketitle
 \section{Introduction}\label{sec:intro}
 The present universe which is isotropic and homogeneous on large scale must have passed through a phase of accelerated exponential expansion called the inflation-in the very beginning of its expansion. Inflation 
 is believed to describe the physics of very early universe quite well and it also explains several conceptual issues in Big-Bang cosmology \cite{Guth, Linde, KT, Baumann}. 
 Most of the ongoing research activities in  the area of cosmic microwave background radiation(CMBR) observations is to converge on the correct model of inflation \cite{Planck2013}. \\ 
%
 Despite of the impressive success of inflation, there are still a couple of pressing issues/questions - related to inflation, period before the commencement of inflation and successful exit from the inflation - are unanswered/unattended till date which might pose few below mentioned alternative possibilities. 
 \begin{itemize}
  \item The universe at its very beginning is described by a quantum gravity theory. It entered into an inflationary phase through the quantum mechanical tunneling.
  \item Prior to the inflation, the universe was dominated by radiation (or some other form of matters e.g. quintom matter, nonlinear electromagnetic field).
  \item The universe underwent a non-singular bounce prior to inflation and subsequently enters into the normal thermal expansion. Prior to the non-singular bounce, the universe was contracting \cite{Wands, Finelli}. 
  \item The universe  existed `eternally' in a quasi-static Minkowski phase of non vanishing minimal radius before it emerges into an inflationary phase and subsequently enters into the normal thermal expansion \cite{Brandenberger, Ellis1, Ellis2}.
 \end{itemize} 
 None of the above listed alternatives is well established till date. A search for the correct alternative (of non-singular cosmology) to Big-Bang cosmology is the need of the hour. 
 \par In this work, we will investigate the emergent universe picture in model of nonlinear electrodynamics within the frame of 4-dimensional Friedmann-Robertson-Walker(FRW) metric.
\noindent    In the early time of the universe evolution
electromagnetic fields were very strong and quantum corrections should be taken into account \cite{Jackson} and, as a result, Maxwell’s electrodynamics becomes nonlinear electrodynamics (NED) \cite{Heisenberg, Schwinger, Adler}. This non-linear electromagnetic fields, coupled strongly to gravity, can induce a non-vanishing trace anomaly term (proportional $T^\mu_\mu$ (trace of the energy-momentum tensor) which is zero for a theory containing only massless fields, and nonzero for a theory containing massive field or in a theory of gravity-NLED fields). The trace anomaly term which can be viewed as a quantum correction to the Einstein-Hilbert action and breaks the scale invariance in the NLED + gravity theory, can generate the negative pressure  \cite{Ali,Kruglov1, Kruglov2, Novello, Ricardo, PS} and hence drive the universe to accelerate. It pushes the universe from its quasi-static Minkowskian phase($-\infty < t \le 0 $) to the inflationary phase of accelerated expansion and subsequently to the normal thermal expansion($t > 0$). 
\par The paper is organised as follows. In Section 2, we describe the general requirements of the emergent universe. In Section 3, we present the analytic and numerical solution of the emergent universe in our model of nonlinear electrodynamics(NLED) with a parameterized EoS parameter $\omega$ near the point of emergence(corresponding to $\omega = -1$). In Section 4, we discuss the slow roll parameters of inflationary expansion in the emergent universe and made an estimate  of the CMB spectral index parameters in this NLED theory. Finally, in Section 5, we summarize our findings and conclude.  

 \section{Emergent Universe-it's requirements}
 Maxwell's electrodynamics usually leads to singular cosmology in various models. It is worthwhile to see whether the non-singular cosmology can be realized in model of nonlinear electrodynamics. \\
 We begin with the spatially flat isotropic and homogeneous universe described by Friedmann-Robertson-Walker(FRW) line element for a flat universe ($k=0$) given by,
 \begin{equation}
  ds^2=dt^2-a^2(t)\left[dr^2 + r^2 (d\theta^2 + sin^2 \theta d\phi^2)\right] 
 \end{equation}
 where $a(t)$ is the scale factor. From the Einstein's equations, for a perfect fluid of density $\rho$ and pressure $p$, we find the energy conservation law, Friedmann equation and Raychaudhuri equation,
\begin{equation}
  \dot{\rho} + 3 H (\rho + P) = 0,
\end{equation}
\begin{equation}  
  ~~~H^2=\left(\frac{\dot{a}}{a}\right)^2 = \frac{\rho}{3}
   \label{FRW1}
\end{equation} 
\begin{equation}  
  ~~~\frac{\ddot{a}}{a}= -\frac{1}{6} (\rho + 3 P)
   \label{FRW2}
\end{equation} 
and taking the time-derivative of the Friedmann, we find
\begin{equation}
  \dot H= \frac{\ddot{a}}{a} - \left(\frac{\dot{a}}{a}\right)^2 = -\frac{1}{2}(\rho+P)
  \label{FRW3}
 \end{equation}
 where, $8\pi G=1$. Here $H = \frac{\dot{a}}{a}$ is the Hubble parameter where $\dot{a}= \frac{da}{dt}$. \\
  The emergent universe is characterised by a quasi-static phase which is followed by an inflationary phase. It corresponds to non-singular cosmology. In this scenario, the universe expands forever starting with a finite scale factor $a(\neq 0)$ at infinite past i.e. its time derivative $\dot a$ is almost zero. The Hubble parameter $H$ , which stays at zero in this static phase, can not be negative. Since in the emergent phase $\rho + P < 0 ~\to \dot{H} > 0$, one finds the EoS parameter($\omega$) in the quasi-static phase 
 \begin{equation}
  \omega= \frac{P}{\rho} = -1-\frac{2~\dot H}{3~H^2} \ll -1
 \end{equation}
 After exiting quasi-Minkowski(steady) state, the universe  enters into thermal expanding phase which suggests $\omega$ should be $-1, 1/3, 0$. This requires a transition from $\omega<-1$ phase to $\omega>-1$ phase. 
%
 It is natural to conclude that for emergent universe scenario, $H^{-1}=\infty$ for $t=-\infty$ which indicates Einstein static(steady) state(ESS) of our universe at infinite past. The evolution of our universe from ESS state requires $H$ should increase and $\dot H>0$ at some point - means the equation of state parameter $\omega<-1$ which violates the null energy condition(NEC) i.e. $\rho + p < 0$. At later time universe enters into an inflationary phase through the time($t_E$) of emergence (at which $\omega = -1$) and  subsequently to normal thermal expansion (radiation and matter dominated phase corresponding to  $\omega=1/3, 0$). 

 \section{Emergent universe with NLED}
 We consider the lagrangian of non-linear electrodynamics that describe the early universe as ,\cite{PS}
 \begin{equation}
  \mathcal{L}=-\frac{\mathcal{F}e^{-\beta\mathcal{F}}}{(\beta\mathcal{F}+1)^2}
  \label{lagrangian1}
 \end{equation}
 where $\mathcal{F}=\frac{B^2-E^2}{2}$, $\beta$ is the non linear parameter having dimension $[M]^{-4}$.  Note that in the limit $\beta\rightarrow 0$ or $\beta \mathcal{F} \to 0$) (in late time cosmology) the lagrangian reduces to classical Maxwell's electrodynamics which is $\mathcal{L} = -\mathcal{F} + \mathcal{O}(\beta \mathcal{F}^2) \to -\mathcal{F} (=  -\frac{1}{4} F_{\mu\nu}F^{\mu\nu})$. So, the Maxwell's theory can be considered as an approximation of weak fields, while for strong fields in early epoch we should use spacetime with NLED fields.\\
The energy-momentum tensor derived for the above lagrangian (Eq.~(\ref{lagrangian1})) can be obtained as  
\begin{equation}
    T_{\mu\nu}=\frac{\partial{\mathcal{L}}}{\partial{\mathcal{F}}}\mathcal{F}_{\mu\alpha}\mathcal{F}^{\alpha}_{\nu}- g_{\mu\nu} \mathcal{L} 
\end{equation}
where $\frac{\partial{\mathcal{L}}}{\partial{\mathcal{F}}} = \frac{e^{- \beta \mathcal{F}}}{(1 + \beta \mathcal{F})^3} \left[ - 1 + 2 \beta \mathcal{F} + \beta^2 \mathcal{F}^2\right]$. 
The trace of the energy-momentum tensor $T^{\mu}_{\mu}$ can be obtained as 
\[T^{\mu}_{\mu} = - 4 \mathcal{L} + \frac{4 e^{- \beta \mathcal{F}}}{(1 + \beta \mathcal{F})^3} \left[ - 1 + 2 \beta \mathcal{F} + \beta^2 \mathcal{F}^2\right] \mathcal{F} \]
which is non-vanishing in the presence of the nonlinear parameter $\beta \neq 0$, however the trace is zero for $\beta = 0$.  So, the scale invariance in the NLED model is broken and this leads us the negative pressure. In the theory where the nonlinear electromagnetic field acts as the source as gravity, to restore the isotropy in Friedman-Robertson-Walker(FRW) space-time, we need to take the average of electromagnetic field. 

According to standard cosmological models a symmetry in the direction holds (i.e. Universe is isotropic). 
Now, the stochastic fluctuations in electron-positron plasma can lead to a stochastic magnetic field \cite{Lemoine1,Lemoine2} which will fill the Universe. 
Assuming that the cosmic background is filled up by the stochastic magnetic field, the averaged magnetic fields which guaranty the isotropy of the Friedman− Robertson− Walker (FRW) space-time, should obey the equations 
\begin{equation}
  <{\bf{E}}>= <{\bf{B}}>=0,~ <E_iB_j>=0,~ <E_iE_j>=\frac{1}{3}E^2\delta_{ij},~  <B_iB_j>=\frac{1}{3}B^2\delta_{ij}
\end{equation}
where $<>$ denotes an average over a volume larger than the the radiation wavelength and smaller as compared with the curvature of space time.\cite{Tolman}\\
We set the electric field $\bf{E} = 0$, as the electric field is screened by the primordial charged plasma, while the magnetic field is not screened and the Universe is called the magnetic Universe. \cite{Lemoine2}.

\par
\begin{figure}[h]
 \centering
  \includegraphics[width=4cm]{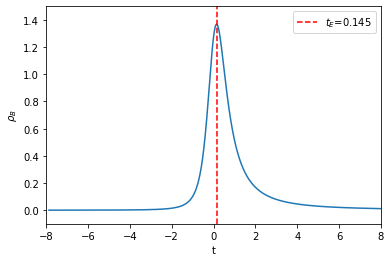}
   \hspace*{0.05in}
   \includegraphics[width=4cm]{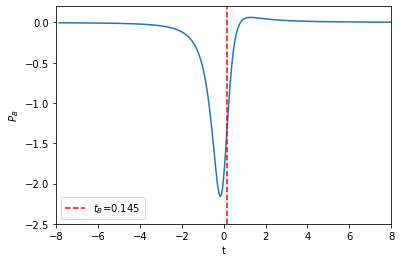}
    \hspace*{0.05in}
   \includegraphics[width=4cm]{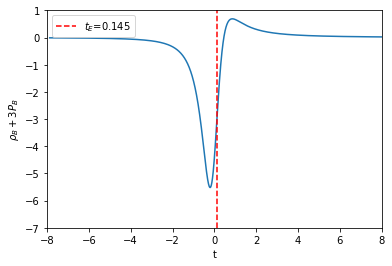}
  \caption{{(Color online) Plot shows the variation of 
  $\rho_B$, $P_B$ and  $\rho_B+3 P_B$ with time $t$. In numerical calculation we took $\beta=0.1$ and $B_0=10^{-10}$ Gauss. The red vertical line in all figures describe the emergent time $t_E =0.145$ s.}}
 \label{Plot2}
  \end{figure}

\begin{figure}[h]
 \centering
   \includegraphics[width=5cm]{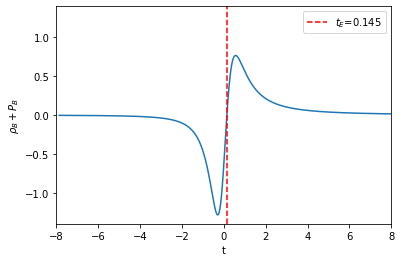}
    \hspace*{0.1in}
  \includegraphics[width=5cm]{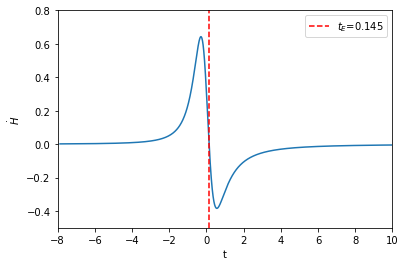}
  \caption{{(Color online) Plot shows the variation of 
  $\rho_B + P_B$ and $\dot{H}$ with time. In numerical calculation we took $\beta=0.1$ and $B_0=10^{-10}$ Gauss.  The red vertical line in all figures describe the emergent time $t_E =0.145$ s.}}
 \label{Plot3}
  \end{figure}

 The energy density($\rho$) and pressure($P$) can be calculated from Eq. (\ref{lagrangian1}) as,
 \begin{equation}
  \rho=-\mathcal{L}-E^2\frac{\partial\mathcal{L}}{\partial\mathcal{F}},~~~
  P=\mathcal{L}-\frac{E^2-B^2}{2}\frac{\partial\mathcal{L}}{\partial\mathcal{F}}
 \end{equation}
For the magnetic universe the energy density $\rho$(=$\rho_B$) and pressure $P$ (=$P_B$) are found to be,
 \begin{equation}
  \rho_B=\frac{2B_0^2}{a^4}e^{-\beta B_0^2/2a^4}\left(2+\frac{\beta B_0^2}{a^4}\right)^{-2}
  \label{rho1}
 \end{equation}
and
\begin{equation}
 P_{B}=-\frac{4B_0^2}{3a^4}\frac{e^{-\beta B_0^2/2a^4}}{\left(2+\frac{\beta B_0^2}{a^4}\right)^3}\left(\frac{\beta^2B_0^4}{a^8}+\frac{11}{2}\frac{\beta B_0^2}{a^4}-1\right)
 \label{p1}
\end{equation}
where we take the magnetic field (following magnetic flux conservation) $B=\frac{B_0}{a^2}$, $B_0=10^{-10}$ G is the present day value of magnetic field\cite{Grasso}.\\
The Hubble parameter can be expressed from Eq.~(\ref{FRW1}) as,
\begin{equation}
 H^2=\frac{2B_0^2}{3a^4}e^{-\beta B_0^2/2a^4}\left(2+\frac{\beta B_0^2}{a^4}\right)^{-2}
 \label{hubble}
\end{equation}
This gives rise to the conservation of energy with effective potential $V_{eff}$ as,
\begin{equation}
 \dot{a}^2 + V_{eff}=0
 \label{veff}
\end{equation}
where $V_{eff}=-\frac{B_0^2}{6a^2}e^{-\beta B_0^2/2a^4}\left(1+\frac{\beta B_0^2}{2a^4}\right)^{-2}$. 
Solving Eq.~(\ref{veff}), we find the evolution of the scale factor $a(t)$ and we express $t$ as a function of $a(t)$ as,
\begin{equation}
 t=\sqrt{\frac{3}{2}}\frac{1}{B_0}\left(a^2e^{\beta B_0^2/4a^4}-\sqrt{\pi\beta}B_0 Erfi\left(\frac{\sqrt{\beta}B_0}{2a^2}\right)\right)
 \label{avst}
\end{equation}
where $Erfi(x)$ is  imaginary Error function and is defined as $Erfi(x)=-iErf(ix)$ where,
$Erf(x)=\frac{2 }{\sqrt{\pi}}\int_0^{x}e^{-t^2}dt$.
\noindent From Eq.~(\ref{rho1}) and Eq.~(\ref{p1}) we can calculate the equation of state parameter as,
\begin{equation}
 \omega(t)=\frac{P_B}{\rho_B}=\frac{2a^8-11\beta B_0^2a^4-2\beta^2B_0^4}{6a^8+3a^4\beta B_0^2}
 \label{omega1}
\end{equation}
where we set  $B_0 = 10^{-10}$ Gauss \cite{Grasso}. 
 \begin{figure}[h]
 \centering
  \includegraphics[width=4cm]{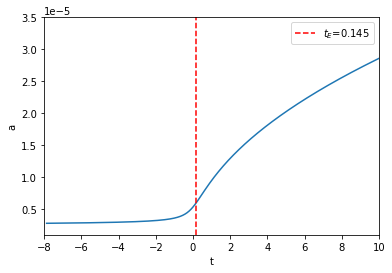}
   \hspace*{0.1in}
  \includegraphics[width=4cm]{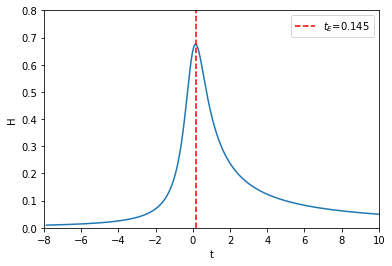}
   \vspace*{-0.25in}
   \includegraphics[width=4cm]{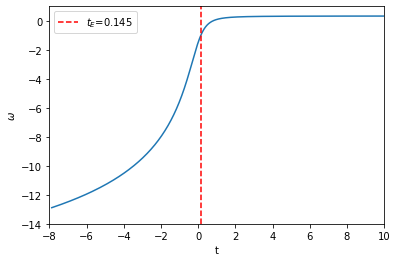}
  \caption{{(Color online) Plot shows the variation of scale factor $a(t)$, hubble parameter $H(t)$ and EOS parameter $\omega(t)$ with time $t$. In numerical calculation we took $\beta=0.1$ and $B_0=10^{-10}$ Gauss. The red vertical line in all figures describe the emergent time $t_E =0.145$ s.}}
 \label{Plot1}
  \end{figure}
\noindent In Fig. ~(\ref{Plot1}), we have plotted $a(t)$ as a function of $t$ (top-left), $H(t)$ against $t$(top-right) and $\omega(t)$ against $t$(lower). From Fig. ~(\ref{Plot1}), we can see that $a$ approaches non-zero constant as $t \to 0$ and stays at that value till the infinite past i.e. $t \to -\infty$. Also, $a$ starts expanding as $t \to 0+$. In Fig.~(\ref{Plot1})(top-right), we see that $H=H_{max}$ at the emergent point $t=t_E$. From the plot of $\omega$ vs $t$(lower plot of Fig.~(\ref{Plot1})), we see that the EoS parameter $\omega(t=t_E) = \omega_E = -1$, where $t = t_E$ is the time of emergence. One finds $t_E$ by solving $\omega_E= -1$ as,
\begin{equation}
 t_E=\sqrt{\frac{3}{2}}\left(\sqrt{\frac{\beta(-1+\sqrt{2})}{2}}e^{1/2(1+\sqrt{2})}-\sqrt{\beta\pi}Erfi(\frac{1}{\sqrt{2(1+\sqrt{2})}})\right)
 \label{tE}
\end{equation}
Note that $t_E$ coincides with $t = 0$ in the case of usual electrodynamics i.e.$t_E = t = 0$ for $\beta = 0$. Also at $t = t_E$, the Hubble parameter reaches its maximum value $H_E = H_{max}$ where 
\begin{equation}
 H_E=\frac{1+\sqrt{2}}{2+\sqrt{2}}\sqrt{\frac{e^{-1/1+\sqrt{2}}}{3\beta(1+\sqrt{2})}}
 \label{hE}
\end{equation} 
and the scale factor $a$ at time $t_E$ is found to be 
\begin{equation}
 a(t=t_E) = a_E=\left(\frac{\beta B_0^2}{2}+\frac{\beta B_0^2}{\sqrt{2}}\right)^{1/4}
\end{equation}
\noindent Using Eq.~(\ref{rho1}) and Eq.~(\ref{p1}) we calculate $\rho_B+P_B$ and $\rho_B+3P_B$ as
\begin{equation}
 \rho_B+P_B=  - 2 \dot{H} = \frac{4B_0^2e^{-\beta B_0^2/2a^4}(4a^8-4\beta B_0^2 a^4-\beta^2B_0^4)}{3(\beta B_0^2+2a^4)^3}
 \label{rhop1}
\end{equation}
and,
\begin{equation}
 \rho_B+3P_B=\frac{4B_0^2e^{-\beta B_0^2/2a^4}(2a^8-5\beta B_0^2 a^4-\beta^2B_0^4)}{(\beta B_0^2+2a^4)^3}
 \label{rho3p1}
\end{equation}
In Fig.~(\ref{Plot2}), we have plotted $\rho_B$(top-left), $P_B$(top-right) and $\rho_B + 3 P_B$(lower) as a function of $t$.
  In the left (of Fig.~(\ref{Plot2})), we see that $\rho_B = 0$  (constant) during the period $-\infty \le t < 0$ and starts deviating from zero at $t = 0$ and becomes maximum $\rho^{max}_B$ at the point of emergence $t = t_E( > 0)$. On the right(top) plot, we see $P_B = - \rho^{max}_B$ at $t = t_E$.  On the lower  plot, we see that $\rho_B + 3 p_B < 0$ for $t \le 0$ which corresponds to the violation of SEC(strong energy condition). At $t = t_E$, we see $\rho_B + 3 P_B (= 2 P_B) \ge 0 $ i.e. strong energy condition(SEC) is satisfied. \\
  \noindent We next study the variation of $\dot{H}$ and $\rho_B + P_B$  with time $t$. Using Eq.~(\ref{hubble}) and Eq.~(\ref{rhop1}), we find 
  \begin{equation}
   \dot H = - \frac{1}{2}(\rho_B + P_B) =\frac{-2B_0^2e^{-\beta B_0^2/2a^4}(4a^8-4\beta B_0^2 a^4-\beta^2B_0^4)}{3(\beta B_0^2+2a^4)^3}
  \end{equation}
  In Fig.~(\ref{Plot3}), we have plotted $\rho_B + P_B$(left)  and $\dot{H}$(right) as a function of $t$.
On the left plot, we see that $\rho_B + p_B <0$ for $t \le 0$ which suggests violation of NEC(null energy condition). 
For $t \ge t_E$, we find $\rho_B + P_B \ge 0$ i.e. NEC is satisfied. On the right plot, we see that $\dot{H} = 0$ at the emergent point $t = t_E$ and $\dot{H}$ moves from $\dot{H} > 0$(NEC condition is violated) region to $\dot{H} < 0$ region (NEC is obeyed) through the point of emergence (i.e. $\dot{H}=0$). 

\begin{figure}[h]
 \centering
   \includegraphics[width=5cm]{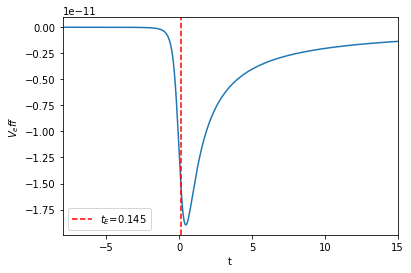}
  \hspace*{0.25in}
  \includegraphics[width=5cm]{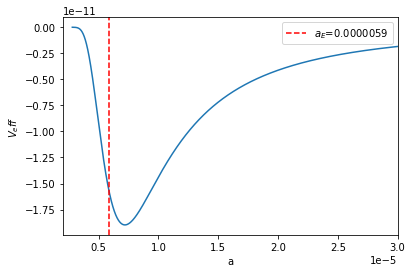}
  \caption{{(Color online) Plot shows the variation of  $V_{eff}$ with time and scale factor $a$ respectively. In numerical calculation we took $\beta=0.1$ and $B_0=10^{-10}$ Gauss.  The red vertical line in all figures describe $a_E$ (the value of $a$ at the emergent time $t_E =0.145$ s).}}
 \label{plot4}
  \end{figure}

\noindent Finally, in Fig.~(\ref{plot4}), we have plotted $V_{eff} $ against $t$(on the left) and $V_{eff} $ against $a$ (on the right) . 
  \noindent We see $V_{eff} = 0$ in the quasi-static Minkowski phase during the period $-\infty < t \le 0$ and it leave this phase at $t \simeq 0$. Note that $V_{eff}$ is negative for all $t \ge 0$ (left plot) and all $a$ (right plot). On the right plot, $V_{eff}$ is minimum at $a=a_c$ (as determined below). On the left of the minimum i.e. $a = a_c$, we find $\frac{dV_{eff}}{da} < 0$, while on the right $\frac{dV_{eff}}{da} > 0$. Noting that   
  \begin{equation}
\dot{a}^2 = - V_{eff} \to 2 \dot{a} \ddot{a} = - \frac{dV_{eff}}{dt} = - \frac{dV_{eff}}{da} \dot{a} \Longrightarrow \ddot{a} = - \frac{1}{2} \frac{dV_{eff}}{da} > 0 ~~\rm{if}~~  \frac{dV_{eff}}{da} < 0 
  \end{equation}
  Again,
\begin{equation}
\ddot{a} = - \frac{1}{2} \frac{dV_{eff}}{da} < 0 ~~\rm{if}~~  \frac{dV_{eff}}{da} > 0 
  \end{equation}  
Now $V_{eff}$ is minimum at $a = a_c$ where 

$a_c = \left(\frac{\beta B_0^2 (5 + \sqrt{33})}{4}\right)^{1/4} = 7.2\times 10^{-6}$
Here we have taken $\beta = 0.1, B_0=10^{-10}$ Gauss. On the left of $a = a_c$(right plot), we see $\frac{dV_{eff}}{da} < 0 ~\to \ddot{a} > 0$ which corresponds to the accelerating universe. \\

\noindent On the left plot, $V_{eff}$ is minimum at $t=t_c$. Now $\frac{dV_{eff}}{dt} < 0$ on the left of the minimum and hence  
  \begin{equation}
\dot{a}^2 = - V_{eff} \to 2 \dot{a} \ddot{a} = - \frac{dV_{eff}}{dt} \Longrightarrow ~~\ddot{a} = - \frac{1}{2 \dot{a}} \frac{dV_{eff}}{dt} > 0~~ \rm{if}~~  \frac{dV_{eff}}{dt} < 0,~\dot{a} > 0 ~(\rm{for}~  t > 0), 
  \end{equation}
i.e. the accelerating universe. The inflationary expansion corresponds to $\dot{\rho_B} = 0 \to \omega = -1$ at the time of emergence $t = t_E$ (where $0 < t_E < t_c$). On the right of the minimum $\frac{dV_{eff}}{dt} > 0$ and we find
\begin{equation}
\ddot{a} = - \frac{1}{2 \dot{a}} \frac{dV_{eff}}{dt} < 0~~ \rm{if}~~  \frac{dV_{eff}}{dt} >  0, ~\dot{a} > 0~(\rm{for}~  t > t_c), 
  \end{equation}  
  the decelerating universe. So, in a nutshell, after exiting the quasi-static Minkowskian space, the universe enters into the inflationary phase of acceleration through the time of emergence $t = t_E$ which corresponds to 
  $\omega = -1$ and $\rho = \rho_{B}^{max} = \rm{constant}$. 
  
  \noindent From Eq.~(\ref{rhop1}) we see that $\rho_B + P_B$ is zero at the emergent time $t = t_E$ with the scale factor  $a = a_E = \left(\frac{\beta B_0^2}{2}+\frac{\beta B_0^2}{\sqrt{2}}\right)^{1/4} = 5.9\times 10^{-6}$ (with $\beta = 0.1, B_0=10^{-10}$ ). We also see that $a_E < a_c$, where $a_c = 7.2\times  10^{-6}$ and $a_E=5.9\times 10^{-6}$ for $\beta = 0.1, B_0=10^{-10}$ Gauss. 
   We also find $t_E = 0.145~\rm{s}$ and $t_c =0.453~\rm{s}$ for $\beta = 0.1$, respectively. 
%

  \subsection{Slow-roll Parameters and CMB constaints}
  The evolution of the scale factor $a$ with time $t$ is shown in Fig.~(\ref{Plot1}) which suggests the usual exponential expansion of $a$ associated with normal slow rolling inflationary model at $t > t_E$(which is $ > 0, \rm{for} ~\beta \neq 0$), while at earlier time $t$ with $t < t_E$,~  $a$ approaches to constant non-zero positive value which corresponds to non-singular cosmology(contrary to the normal singular big-bang cosmology). 
  
  \noindent The duration of slow roll regime is described by two slow-roll parameters \cite{Baumann},
  \begin{equation}
   \epsilon=-\frac{\dot H}{H^2} ,  ~
   \eta=\epsilon-\frac{1}{2\epsilon}\frac{d\epsilon}{dN}
   \label{parameters}
  \end{equation}
  Where $N$ is the e-fold number that can be expressed as, $dN=H dt$.  Slow roll approximation requires $\epsilon<<1$ and $\eta<<1$.  
For NLED case, Eq.~(\ref{parameters}) can be expressed in terms of magnetic field as\cite{PS},
\begin{equation}
 \epsilon=2B\frac{H_{,B}}{H} , ~\eta=\epsilon-\frac{a}{2\epsilon}\frac{d\epsilon}{da}
\end{equation}
From Eq.~(\ref{hubble}), we can express the slow roll parameters $\epsilon$ and $\eta$ for emergent universe as,
\begin{equation}
 \epsilon=\frac{-\beta^2 B_0^4-4\beta B_0^2 a^4+4a^8}{a^4(\beta B_0^2+2a^4)}
 \label{epsilon}
\end{equation}
and,
\begin{equation}
 \eta=\frac{16a^{16}-56\beta B_0^2 a^{12}+6\beta^3 B_0^6a^4+\beta^4B_0^8}{a^4(-\beta^3B_0^6-6\beta^2B_0^4a^4-4\beta B_0^2 a^8+8a^{12})}
 \label{eta}
\end{equation}
Next, we demostrate that emergent universe described by Eq.~(\ref{lagrangian1}) is consistent with recent measurement of Cosmic Microwave Background (CMB) by Planck 2018 \cite{Planck2018}. As seen from Fig.~(\ref{plot4}), $V_{eff}$ vs $t$ plot, universe commences towards Inflation after leaving quasi-Minkowski state. Quantum fluctuations generated during Inflation can explain density perturbations observed in CMB. Those primordial density fluctuations can be characterised by scalar spectral index $n_s$, tensorial spectral index $n_T$ and tensor to scalar ratio $r$ and in slow-roll approximation,they are related as
\begin{equation}
 r=4(1-n_s)=-8n_T
\end{equation}
In slow roll approximation, $n_s$, $r$ and $n_T$ can be expressed as,
\begin{equation}
 n_s-1=2\eta-4\epsilon, ~ r=16\epsilon, ~ n_T=-2\epsilon
 \label{ns}
\end{equation}

\noindent From Eq.~(\ref{ns}) and Eq.~(\ref{epsilon}) we can derive $r$ as
\begin{equation}
 r=16\frac{-\beta^2 B_0^4-4\beta B_0^2 a^4+4a^8}{a^4(\beta B_0^2+2a^4)}
\end{equation}
and the scalar spectral index $n_s$ as 
\begin{equation}
 n_s=1+\frac{4(\beta^2B_0^4+4\beta B_0^2 a^4-4a^8)}{a^4(\beta B_0^2+2a^4)}
\end{equation}

Below in Table 1, we have presented our predicted values of $r$, $n_s$ and $n_T$ corresponding to different $\beta$ values. Also shown in the table are the emergent time $t_E$, the scale factor $a_E$ (at time $t_E$) and the non-linear parameter $\beta$  corresponding to the present day magnetic field $B_0 = 10^{-10}$ G \cite{Grasso}.

\begin{table}[!ht]
\centering

\begin{tabular}{||c c c c c c c c||}
\hline
  $\beta$ & $B_0$ & $t_E$ & $a_E$ & $n_s$ & $r$ & $n_T$ & \\ [0.5ex]
 
\hline\hline
 $0.1$ & $10^{-10}$ & 0.1450 & $5.90\times 10^{-6}$ & $0.97467$ & $0.10133$ & $-0.012670$  & \\
 
 $0.01$ & $10^{-10}$ & $0.0458$ & $3.32\times 10^{-6}$ & $0.95727$ & $0.17093$ & $-0.021367$ & \\
 
 $0.001$ & $10^{-10}$ & 0.0145 & $1.86\times 10^{-6}$ & $0.99942$ & $0.00231$ & $-0.000289$ & \\ 
 \hline
 \end{tabular}
 \caption{ Note that the scalar spectral index $n_s = 0.97 \pm 0.02$ (from SDSS-III/BOSS DR9 data \cite{Delabrouille}) and $n_s=0.9649\pm0.0042$ from PLANCK 2018 Data \cite{Planck2018}.}
\label{table:1}
\end{table}
From the Table \ref{table:1}, we see that the $n_s = 0.97467(0.99942)$ corresponding to $\beta = 0.1(0.001)$ lies within within $1 \sigma(2 \sigma)$ of the SDSS-III/BOSS DR9 data  and within $ 3~\sigma$ of the Planck data corresponding to $\beta = 0.1$. We also find the tensor-to-scalar ratio $r \approx 0.10$ and tensor spectral index $n_T (= -r/8) = -0.012670(-0.000289)$ corresponding to $\beta = 0.1(0.001)$ which is close to the Planck estimate.  
\section{Conclusion}
We study the emergent universe (non-singular universe at time $t=0$) in a class of models of nonlinear electrodynamics which is characterised by a dimensionful parameter $\beta$.  The scale factor $a(t)$ in this NLED scenerio is found to remain constant during the phase of infinite past i.e. $-\infty < t \le 0$ (the phase is called the quasi-static Minkowski phase) and remains so till time $t = 0$. The non-zero NLED parameter $\beta$ allows the universe from its quasi-static Minkowski phase to enter into the inflationary phase of expansion at the time of emergence $t_E$ at which the EoS parameter $\omega = \omega_E = -1$ and subsequently to the normal expanding phase dominated by dust matter. We find that the Hubble parameter($H$) is maximum $H=H_{max}$ and the Null Energy Condition(NEC) is violated at the time of emergence $t_E$ which is found to be dependent on $\beta$. We found the time of emergence $t = t_E = 0.145$ s (which corresponds to scale factor $a_E = 5.9\times 10^{-6}$) for $\beta = 0.1$. 
 We estimate the scalar spectral index $n_s = 0.97467$ corresponding to $\beta = 0.1(0.001)$ which lies within within $1 \sigma(2 \sigma)$ of the SDSS-III/BOSS DR9 data and within $ 3~\sigma$ of the Planck data corresponding to $\beta = 0.1$. We also find the tensor-to-scalar ratio $r \approx 0.10$ and tensor spectral index $n_T (= -r/8) = -0.012670(-0.000289)$ corresponding to $\beta = 0.1(0.001)$ which can be compared with the estimate of the Planck 2018 collaboration.
\section{Aknowledgement}
PS would like to thank Department of Science and Technology, Government of India for INSPIRE fellowship. The work of PKD is supported by CSIR Grant No.25(0260)/17/EMR-II. 

 \end{document}